# アプリケーション自動オフロードにおけるリソース量設定の検討

山登庸次†

† NTT ネットワークサービスシステム研究所，東京都武蔵野市緑町 3-9-11
E-mail: †yoji.yamato.wa@hco.ntt.co.jp

**あらまし** 近年，少コアの CPU だけでなく，GPU，FPGA，メニーコア CPU 等のヘテロなデバイスを利用したシステムが増えている．しかし，これらの利用には，OpenMP，CUDA や OpenCL 等のハードウェアを意識した技術仕様の理解が必要であり，ハードルは高い．これらの背景から，私は，プログラマーが CPU 向けに開発したソースコードを，適用される環境に応じて，自動で変換し，リソース量等を設定して，高い性能で運用可能とする環境適応ソフトウェアのコンセプトを提案している．しかし，従来，移行先環境に合わせコードを変換することは検討されているが，リソース量を適切に設定する研究は無かった．本稿では，環境適応ソフトウェアの新たな要素として，アプリケーションをコストパフォーマンス高く動作させるため，CPU とオフロードデバイスのリソース量を適切化する手法を検討する．

**キーワード** 環境適応ソフトウェア, GPGPU，自動オフロード，進化的計算，リソース量設定

# Study of Resource Amount Configuration for Automatic Application Offloading

Yoji YAMATO†

† Network Service Systems Laboratories, NTT Corporation, 3-9-11, Midori-cho, Musashino-shi, Tokyo
E-mail: †yoji.yamato.wa@hco.ntt.co.jp

**Abstract** In recent years, utilization of heterogeneous hardware other than small core CPU such as GPU, FPGA or many core CPU is increasing. However, when using heterogeneous hardware, barriers of technical skills such as OpenMP, CUDA and OpenCL are high. Based on that, I have proposed environment-adaptive software that enables automatic conversion, configuration, and high performance operation of once written code, according to the hardware to be placed. However, although the conversion of the code according to the migration destination environment has been studied so far, there has been no research to properly set the resource amount. In this paper, as a new element of environment adaptive software, in order to operate the application with high cost performance, I study a method to optimize the resource amount of CPUs and offload devices.

**Key words** Environment Adaptive Software, GPGPU, Automatic Offloading, Evolutionary Computation, Resource Amount Configuration

## 1. はじめに

近年，CPU の半導体集積度が 1.5 年で 2 倍になるというムーアの法則が減速するのではないかと言われている．そのような状況から，CPU だけでなく，FPGA（Field Programmable Gate Array）や GPU（Graphics Processing Unit）等のデバイスの活用が増えている．例えば，Microsoft 社は FPGA を使って Bing の検索効率を高めるといった取り組みをしており [1]，Amazon 社は，FPGA, GPU 等をクラウドシステム（例えば，[2]- [8]）のインスタンスとして提供している [9]．また，センサ等のリソースが少ない IoT デバイスを用いた IoT システム（例えば，[10]- [14]）も，サービス連携技術等（例えば，[15]- [24]）を用いて開発され増えてきている．

しかし，少コアの CPU 以外のデバイスをシステムで適切に活用するためには，デバイス特性を意識した設定やプログラム作成が必要であり，OpenMP（Open Multi-Processing）[25]，OpenCL（Open Computing Language）[26]，CUDA（Compute Unified Device Architecture）[27] やアセンブリといった知識が必要になってくるため，大半のプログラマーにとっては，スキルの壁が高い．

少コアの CPU 以外の GPU や FPGA，メニーコア CPU 等のデバイスを活用するシステムは今後ますます増えていくと予想されるが，それらを最大限活用するには，技術的壁が高い．そこで，そのような壁を取り払い，少コアの CPU 以外のデバイスを十分利用できるようにするため，プログラマーが処理ロジックを記述したソフトウェアを，配置先の環境（FPGA，GPU，メニーコア CPU 等）にあわせて，適応的に変換，設定し，環境に適合した動作をさせるような，プラットフォームが求められている．

Java [28] は 1995 年に登場し，一度記述したコードを，別メーカーの CPU を備える機器でも動作可能にし，環境適応に関するパラダイムシフトをソフト開発現場に起こした．しかし，移行先での性能については，適切であるとは限らなかった．そこで，私は，一度記述したコードを，配置先の環境に存在する GPU や FPGA，メニーコア CPU 等を利用できるように，変換，リソース設定等を自動で行い，アプリケーションを高性能に動作させることを目的とした，環境適応ソフトウェアを提案した．合わせて，環境適応ソフトウェアの要素として，アプリケーションコードのループ文及び機能ブロックを，FPGA，GPU に自動オフロードする方式を提案評価している [29]．

本稿は，通常の CPU 向けプログラムを，GPU 等のデバイスにオフロードした際に，アプリケーションをコストパフォーマンス高く動作させるため，CPU とオフロードデバイスのリソース量を適切化する手法を検討する．

## 2. 既 存 技 術

### 2.1 市 中 技 術

環境適応ソフトウェアとしては，Java がある．Java は，仮想実行環境である Java Virtual Machine により，一度記述した Java コードを再度のコンパイル不要で，異なるメーカー，異なる OS の CPU マシンで動作させている（Write Once, Run Anywhere）．しかしながら，移行先で，どの程度性能が出るかはわからず，移行先でのデバッグや性能に関するチューニングの稼働が大きい課題があった（Write Once, Debug Everywhere）．

GPU の並列計算パワーを画像処理でないものにも使う GPGPU（General Purpose GPU）を行うための環境として CUDA が普及している．CUDA は GPGPU 向けの NVIDIA 社の環境だが，FPGA，メニーコア CPU，GPU 等のヘテロなデバイスを同じように扱うための仕様として OpenCL が出ており，その開発環境 [30] [31] も出てきている．CUDA，OpenCL は，C 言語の拡張を行いプログラムを行う形だが，プログラムの難度は高い（FPGA 等のカーネルと CPU のホストとの間のメモリデータのコピーや解放の記述を明示的に行う等）

CUDA や OpenCL に比べて，より簡易にヘテロなデバイスを利用するため，指示行ベースで，並列処理等を行う箇所を指定して，指示行に従ってコンパイラが，GPU，メニーコア CPU 等に向けて実行ファイルを作成する技術がある．仕様としては，OpenACC [32] や OpenMP 等，コンパイラとして PGI コンパイラ [33] や gcc 等がある．

CUDA，OpenCL，OpenACC，OpenMP 等の技術仕様を用いることで，FPGA や GPU，メニーコア CPU へオフロードすることは可能になっている．しかしデバイス処理自体は行えるようになっても，高速化することには課題がある．例えば，マルチコア CPU 向けに自動並列化機能を持つコンパイラとして，Intel コンパイラ [34] 等がある．これらは，自動並列化時に，コードの中のループ文中で並列処理可能な部分を抽出して，並列化している．しかし，メモリ処理等の影響で単に並列化可能ループ文を並列化しても性能がでないことも多い．FPGA や GPU 等で高速化する際には，OpenCL や CUDA の技術者がチューニングを繰り返したり，OpenACC コンパイラ等を用いて適切な並列処理範囲を探索し試行することがされている．

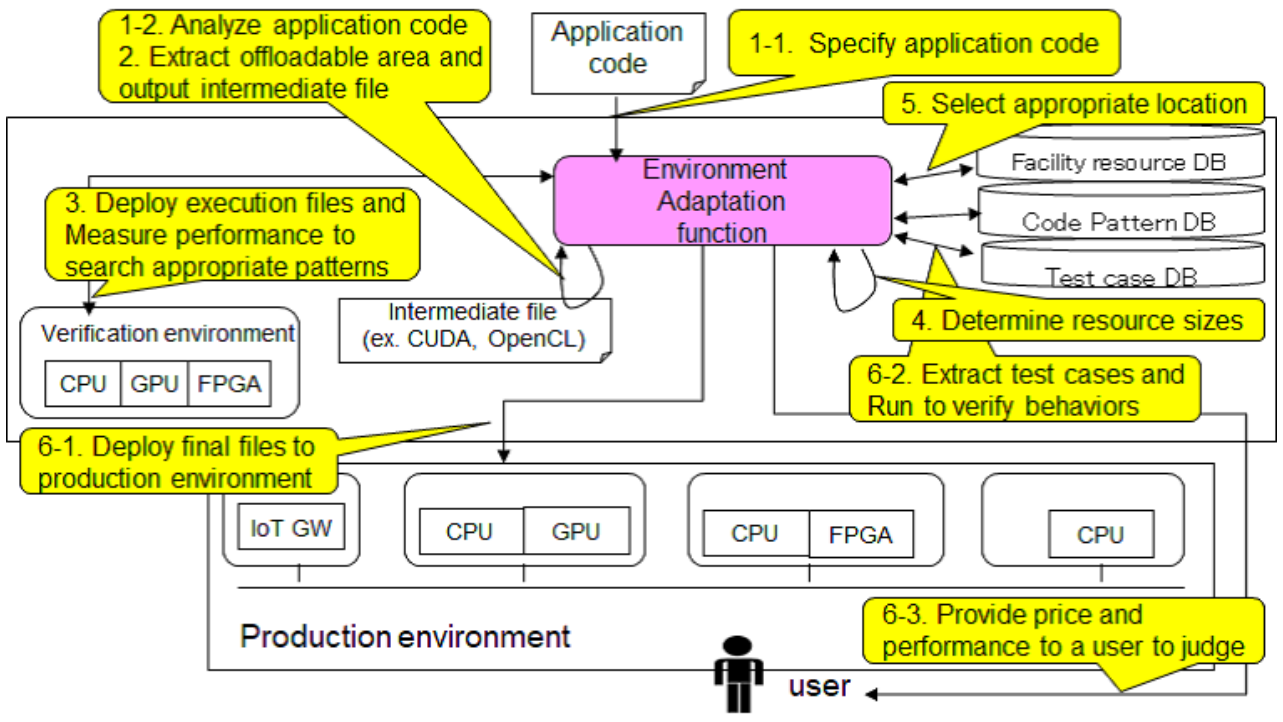

図 1　環境適応ソフトウェアのフロー

このため，技術スキルが乏しいプログラマーが，FPGA や GPU，メニーコア CPU を活用してアプリケーションを高速化することは難しいし，自動並列化技術等を使う場合も並列処理箇所探索の試行錯誤等の稼働が必要だった．並列処理箇所探索の試行錯誤を自動化する取り組みとして，著者は進化計算手法を用いた GPU 自動オフロードを提案している．

### 2.2 環境適応ソフトウェア提案

ソフトウェアの環境適応を実現するため，著者は図の処理フローを提案している．環境適応ソフトウェアは，環境適応機能を中心に，検証環境，商用環境，テストケース DB，コードパターン DB，設備リソース DB の機能群が連携することで動作する．

Step1 コード分析：

Step2 オフロード可能部抽出：

Step3 適切なオフロード部探索：

Step4 リソース量調整：

Step5 配置場所調整：

Step6 実行ファイル配置と動作検証：

Step7 運用中再構成：

ここで，Step 1-7 で，環境適応するために必要となる，コードの変換，リソース量の調整，配置場所の決定，検証，運用中の再構成を行うが，実施したい処理だけ切り出すこともできる．例えば，GPU 向けのコード変換だけ実施する場合は，Step 1-3 だけ処理すればよい．

GPU 向けにループ文を自動オフロードする例として，著者は [29] を提案している．提案方式では，GPU 処理に適切なループ文を，検証環境での繰返し測定により抽出を行うが，その際

に進化計算手法の一つである遺伝的アルゴリズム [35] を用いて探索を行っている．さらに，CPU と GPU 間のメモリデータ転送を削減するため，ネストループ内の変数で一括化できる変数は一括転送を行っている．

現状を整理する．ヘテロなデバイスに対するオフロードは手動での取組みが主流である．著者は環境適応ソフトウェアのコンセプトを提案し，自動オフロードを検討しているが，自動オフロードした後の，オフロードデバイスのリソース量については検討がされていない．そのため，本稿では，GPU 等のオフロードデバイスに自動オフロードした際の，CPU とオフロードデバイスのリソース量の適切化を対象とする．

## 3. CPU とオフロードデバイスのリソース量適切化

著者は，環境適応ソフトウェアのコンセプトを具体化するために，これまでに，プログラムのループ文の GPU 自動オフロード，FPGA 自動オフロード，プログラムの機能ブロックの自動オフロードの方式を提案してきた（[29], [36]- [39]）．これらの要素技術検討も踏まえて，本節の，3.1 では，CPU とオフロードデバイスのリソース比を適切にするための検討する，3.2 では，リソース量決定とその検証について検討する．

### 3.1 CPU とオフロードデバイスのリソース比適切化

[29] 等の手法により，GPU や FPGA 等のオフロードデバイスに通常のプログラムを自動オフロードすることができる．本サブ節では，デバイスにオフロードするプログラム変換が出来た後の，CPU とオフロードデバイスのリソース比の適切化について検討する．

現在，マルチコア CPU，メニーコア CPU は，仮想マシンやコンテナによる仮想化により，全コアの何割を割り当てる等が柔軟にできるようになっている．GPU についても，近年 CPU 同様の仮想化がされるようになってきており，GPU の全コアの何割を割り当てる等の運用が可能になりつつある．FPGA に関しては，リソース使用量は，Look Up Table や Flip Flop の設定数で表されることが多く，利用されていないゲートについては別用途に使うことができる．このように，CPU，GPU，FPGA とも全リソースの一部を使う運用が可能であり，CPU とオフロードデバイスのリソースを用途に応じて適切化することはコストパフォーマンスを高める上で重要である．

アプリケーションを CPU と GPU 処理のコードに [29] 等の手法で変換し，コード自体は適切であっても，CPU と GPU とのリソース量が適切なバランスでない場合は，性能が出ない．例えば，ある処理を行う際に，CPU の処理時間が 1000 秒，GPU の処理時間が 1 秒では，オフロードできる処理を GPU である程度高速化しても，全体的には CPU がボトルネックとなっている．[40] では，CPU と GPU を使って MapReduce フレームワークでタスク処理している際に，CPU と GPU の実行時間が同じになるよう Map タスクを配分することで，全体の高性能化を図っている．

本稿では，CPU とオフロードデバイスのリソース比を決めため，[40] 等も参考に，何れかのデバイスでの処理がボトルネックとなる事を避けるため，テストケースの処理時間から，CPU とオフロードデバイスの処理時間が同等オーダーになるように，リソース比を決定することを提案する．

[29] の手法も該当するが，著者は自動オフロードの際は，検証環境での性能測定結果に基づいて徐々に高速化していく手法を提案している．理由としては，性能に関しては，コード構造だけでなく，実際に処理するハードウェアのスペック，データサイズ，ループ回数等の実際に処理する内容によって大きく変わるため，静的に予測する事が困難であり，動的な測定が必要だからである．そのため，コード変換の際に，既に検証環境での性能測定結果があるので，その結果を用いてリソース比を定める．

性能測定の際には，テストケースを指定して測定を行うが，例えば，検証環境でのテストケースの処理時間が，CPU 処理：10 秒，GPU 処理：5 秒の場合では，CPU 側のリソースは 2 倍で同等の処理時間程度と考えられるため，リソース比は 2:1 となる．なお，特にある処理をオフロードで高速化したいといったユーザ要望は，その処理を含むテストケースを準備して，そのテストケースに対して [29] 等の手法で高速化することで反映される．

### 3.2 CPU とオフロードデバイスのリソース量決定と自動検証

前サブ節で，リソース比が定まったため，次に商用環境へのアプリケーションの配置を行う．商用環境への配置の際は，ユーザが指定したコスト要求を満たすように，リソース比は可能な限りキープして，リソース量を決定する．例えば，CPU に関して 1VM は 1000 円/月，GPU は 4000 円/月，リソース比は 2:1 が適切で，ユーザは月 10000 円以内の予算だった場合には，CPU は 2，GPU は 1 を確保して商用環境に配置する．また，月 5000 円以内の予算だった場合には，リソース比はキープできないが，CPU は 1，GPU は 1 を確保して配置する．

商用環境にリソースを確保してプログラムを配置した後は，ユーザが利用する前に動作することを確認するため，自動検証が行われる．自動検証では，性能検証テストケースやリグレッションテストケースが実行される．性能検証テストケースは，ユーザが指定した想定テストケースを Jenkins 等の試験自動実行ツールを用いて行い，処理時間やスループット等を測定する．リグレッションテストケースは，システムにインストールされるミドルウェアや OS 等のソフトウェアの情報を取得して，それらに対応するリグレッションテストを Jenkins 等を用いて実行する．これらの自動検証を，少ないテストケースの準備で行うための検討は [41] 等でされており，それを用いる．

性能検証テストケースでは，オフロードした場合でも計算結果が不正でないかを，オフロードしない場合との計算結果差分のチェックもする．例えば，GPU を処理する PGI コンパイラは，PCAST という機能の pgi_compare や acc_compare という API で，GPU を使う場合使わない場合の計算結果差分を確認できる．なお，GPU と CPU では丸め誤差が異なる等，並列処理等を正しくオフロードしても完全に計算結果が一致しない場合もある．そのため，IEEE 754 仕様での確認等を行い，許

容できる差分かをユーザに提示して確認頂く．

自動検証の結果として，性能検証テストケースの処理時間やスループットや計算結果差分，リグレッションテストの実行結果の情報が，ユーザに提示される．ユーザには合わせて，確保したリソース（VM の数やスペック等）とその価格が提示されており，それら情報を参照してユーザは運用開始を判断する．

## 4. 実　　装

### 4.1 利用ツール

提案技術の有効性を確認するため現在実装している実装の設計を説明する．対象アプリケーションは C/C++言語のアプリケーションとし，GPU オフロード時のリソース適切化を確認する．GPU は，GeForce RTX 2080 Ti (CUDA core: 4352, Memory：GDDR6 11GB) を用いる．

GPU 処理は市中の PGI コンパイラ 19.10 を用いる．PGI コンパイラは OpenACC を解釈する C/C++/Fortran 向けコンパイラであり，for 文等のループ文を，OpenACC のディレクティブ #pragma acc kernels, #pragma acc parallel loop で指定することにより，GPU 向けバイトコードを生成し，実行により GPU オフロードを可能としている．

CPU の仮想化及び GPU の仮想化は Citrix Xen Server 4 を用いて仮想化を行い，CPU や GPU のリソースの柔軟な分割，割り当てを行う．

C/C++言語の構文解析には，LLVM/Clang 6.0 の構文解析ライブラリ (libClang の python binding) を用いる．

実装は Perl 5 と Python 2.7 で行い，以下の処理を行う．Perl は GA の処理を中心に，Python は構文解析等のその他の処理を中心に行う．

### 4.2 実装動作

実装の動作概要を示す．実装は，ユーザからアプリケーションのオフロード依頼があると，構文解析ライブラリを用いてコード解析を行う．次に，ループ文オフロードの試行を行い，検証環境での性能測定を通じて高性能のパターンを解とする．解のパターンに対する性能測定結果を元に，CPU と GPU の適切なリソース比を算出する．適切なリソース比と，ユーザの依頼の条件から，リソース量を定める．

#### 4.2.1 ループ文オフロード試行

実装は，まず，C/C++アプリケーションのコードを Clang で解析して，for 文を発見するとともに，for 文内で使われる変数データ，その変数の処理等の，プログラム構造を把握する．

並列処理自体が不可な for 文は排除する必要がある．各 for 文に対して，GPU で処理するディレクティブ挿入を試行し，エラーが出る for 文は GA の対象外とする．ここで，エラーが出ないループ文の数が a の場合，a が遺伝子長となる．

次に，初期値として，指定個体数の遺伝子配列を準備する．遺伝子の各値は，0 と 1 をランダムに割当てて作成する．準備された遺伝子配列に応じて，遺伝子の値が 1 の場合は GPU 処理を指定するディレクティブを C/C++コードに挿入する．

次に，変数データの参照関係を元に，GPU 向けデータ転送を指示する．転送が必要なケースは，CPU プログラム側で設定，定義した変数と GPU プログラム側で参照する変数が重なる場合は，CPU から GPU への変数転送が必要であり，GPU プログラム側で設定した変数と CPU プログラム側で参照，設定，定義する変数が重なる場合は，GPU から CPU への変数転送が必要である．転送必要な変数について，GPU 処理開始前と終了後に一括転送すればよい変数については，複数ファイルで定義された変数を一括転送するディレクティブを挿入する．

ディレクティブを挿入された C/C++コードを，PGI コンパイラでコンパイルを行う．コンパイルした実行ファイルをデプロイし性能を測定する．性能測定では処理時間とともに，例えば，PGI コンパイラの PCAST 機能を用いて並列処理した場合の計算結果が，元のコードと大きく差分がないかチェックし，許容外の場合は，処理時間を∞とする．

全個体に対して，性能測定後，処理時間に応じて，各個体の適合度を設定する．設定された適合度に応じて，残す個体の選択を行う．選択された個体に対して，交叉処理，突然変異処理，そのままコピー処理の GA 処理を行い，次世代の個体群を作成する．

次世代の個体に対して，指示挿入，コンパイル，性能測定，適合度設定，選択，交叉，突然変異処理を行う．指定世代数の GA 処理終了後，最高性能の遺伝子配列に該当する，ディレクティブ付き C/C++コードを解とする．

#### 4.2.2 リソース比とリソース量の決定

リソース比を適切化するため，オフロードパターンの解を決める際の性能測定結果を用いる．実装は，テストケースの処理時間から，CPU と GPU の処理時間が同等オーダーになるようリソース比を定める．例えば，テストケースの処理時間が，CPU 処理：10 秒，GPU 処理：5 秒の場合では，CPU 側のリソースは 2 倍で同等の処理時間程度と考えられるため，リソース比は 2:1 となる．なお，仮想マシン等の数は整数となるため，リソース比は処理時間から計算する際に，整数比となるように四捨五入をする．

リソース比が定まると，次に，商用環境へのアプリケーション配置を行う際のリソース量決定に入る．実装は，リソース量決定には，ユーザがオフロード依頼時に指定したコスト要求を満たすように，リソース比は出来るだけキープして，VM 等の数を定める．具体的には，コスト範囲内で，リソース比をキープする中では，VM 等の数は最大値を選択する．例えば，CPU に関して 1VM は 1000 円/月，GPU は 4000 円/月，リソース比は 2:1 が適切で，ユーザは月 10000 円以内の予算だった場合には，CPU は 2，GPU は 1 を確保する．コスト範囲内で，リソース比をキープできない場合は，CPU1 単位，GPU1 単位から始めて出来るだけ適切なリソース比に近くなるよう，リソース量を定める．例えば，月 5000 円以内の予算だった場合には，リソース比はキープできないが，CPU は 1，GPU は 1 を確保する．リソース量が決まったら，実装は，Xen Server の機能を用いて，CPU や GPU のリソースを割り当てる．

## 5. 関連研究

OpenStack を用いたクラウド上で，リソース効率化を行う研

究として，[42] がある．本研究も，クラウド含めたネットワークワイドでのリソース効率化技術と言えるが，本研究は，特にヘテロデバイスが提供される際に，適切にオフロードデバイスのリソース量を設定するための技術である．[43] は CPU と GPU が同じダイの統合チップの際の，タスクのスケジューリングに関するもので，CPU と GPU にどの程度タスクを割り当てるかを参考にできる．

GPU へのオフロードについては，[44] [45] [46] 等があげられる．[44] は，C++ expression template の GPU オフロードのため，メタプログラミングと JIT コンパイル利用をあげており，[45] [46] は OpenMP 使った GPU へのオフロードに取り組んでいる．新たな開発モデルや指示句の手動挿入等を行わず，著者が狙うような既存コードを自動で GPU 向けに変換する研究は少ないと言える．

自動で GPU にオフロードする領域を探索する技術として，[47] が上げられる．[47] は，GA によりオフロードする部分の最適化を行っている点で，著者の以前研究と同様である．しかし，[47] は，流体計算の姫野ベンチマーク等，GPU での高速化が数多く行われているアプリケーションを対象としており，母集団 20，世代数 200 で評価を行い，長時間の繰返し実行を必要としている．今回も利用している著者の以前提案技術は，CPU 向け汎用アプリケーションを GPU で高速化する際に，一定時間で利用開始できることを狙っている．

今回検証では，C/C++言語の OpenACC を解釈する PGI コンパイラを用いて検証を行った．OSS で良く使われる言語として，Java と Python がある．Java では，Java 8 より lambda 形式での並列処理記述が可能である．IBM は，lambda 形式の並列処理記述を，GPU にオフロードする JIT コンパイラを提供している [48]．Java では，これらを用いて，ループ処理を lambda 形式にするか否かのチューニングを GA で行うことで，同様のオフロードが可能である．Python では PyACC というサンプル実装があり，Python で OpenACC を解釈できるため，C 言語同様の自動オフロードが可能と考える．

OpenCL, CUDA は基本的にノード内での並列処理を制御するが，マルチノードで処理を行う場合は，MPI によりノード間並列処理を行うことが一般的である．しかし，MPI も OpenCL, CUDA 同様，並列処理の高い技術知識が要求される．そこで，MPI を隠蔽し，別ノードのデバイスもローカルノードのデバイスとして見せ，別ノードのデバイスを制御する MPI 隠蔽技術も出てきている [49]．マルチノードにも検討を拡張する際は，これら MPI 隠蔽技術も利用する．

以上の様に，GPU，FPGA，メニーコア CPU へのオフロードによる高速化は数多くの取り組みがあるが，OpenMP のようにどの部分を並列化するといった指示を手動で追加し，それに従ってオフロードする取組みが主流であり，既存コードを自動でオフロードするような取組みはほとんどない．また，GPU，FPGA，メニーコア CPU へオフロードするための変換の検討だけであり，本稿で対象としているような，GPU 等のオフロードデバイスのリソース量の適切化による，コストパフォーマンスを高める検討はない．

## 6. ま と め

本稿では，私が提案している，ソフトウェアを配置先環境に合わせて自動適応させ GPU，FPGA，メニーコア CPU 等を適切に利用して，アプリケーションを高性能に運用するための環境適応ソフトウェアの要素として，GPU 等のデバイスに自動オフロードした際に，コストパフォーマンスを高めるため，オフロードデバイスのリソース量を適切化する手法を提案した．

GPU 等のデバイスで処理できるよう，プログラムを変換し，自動オフロードされた後に提案方式は動作する．提案方式は，まず，CPU とオフロードデバイスのリソース比を適切化するため，何れかのデバイスでの処理がボトルネックとなる事を避けるよう，検証環境でのテストケースの処理時間から，CPU とオフロードデバイスの処理時間が同等オーダーになるように，リソース比を決定する．提案方式は，次に，そのリソース比に基づいて，商用環境への配置を行うが，ユーザが指定したコスト要求を満たすように，適切リソース比は可能な限りキープして，リソース量を決定する．提案方式は，プログラムとリソース配置後，ユーザが利用する前に動作することを確認するため，自動検証を行い，性能，オフロード計算妥当性，正常動作結果，リソース価格をユーザに提供し，開始判断してもらう．

今後は，提案方式の実装を進め市中アプリケーションで有効性を確認する．更に，オフロードデバイスの配置場所をクラウドにするかエッジにするか等の配置場所の適切化検討を行う．

## 文 献